\documentclass[a4paper,11pt]{article}
\pdfoutput=1 % if your are submitting a pdflatex (i.e. if you have
\usepackage{jcappub} % for details on the use of the package, please
\usepackage[T1]{fontenc} % if needed

\newcommand{\tr}[1]{\textcolor{black}{#1}}

\title{\boldmath Neutrino physics with multi-ton scale liquid xenon detectors}

\author[a,1]{L. Baudis,\note{Corresponding author.}}
\author[a,b]{A. Ferella,}
\author[a]{A. Kish,}
\author[a,e]{A.~Manalaysay,}
\author[a,c]{T.~Marrod\'an Undagoitia}
\author[a,d]{and M. Schumann}

\affiliation[a]{Physik Institut, University of Z\"urich, Winterthurerstrasse 190, CH-8057 Z\"urich, Switzerland}
\affiliation[b]{Laboratorio Nazionale del Gran Sasso, 67010 Assergi, Italy}
\affiliation[c]{Max-Planck-Institut f\"{u}r Kernphysik, 69117 Heidelberg, Germany}
\affiliation[d]{Albert Einstein Center for Fundamental Physics, Universit\"{a}t Bern, 3012 Bern, Switzerland}
\affiliation[e]{Department of Physics, University of California Davis, 95616 Davis, CA, USA}

\emailAdd{laura.baudis@physik.uzh.ch}
\emailAdd{alfredo.ferella@lngs.infn.it}
\emailAdd{alexkish@physik.uzh.ch}
\emailAdd{aaronm@ucdavis.edu}
\emailAdd{marrodan@mpi-hd.mpg.de}
\emailAdd{marc.schumann@lhep.unibe.ch}

\abstract{We study the sensitivity of large-scale xenon detectors to low-energy solar neutrinos, to coherent neutrino-nucleus scattering and to neutrinoless double beta decay.
As a concrete example, we consider the xenon part of the proposed {\sc DARWIN} (Dark Matter WIMP Search with Noble Liquids) experiment.
We perform detailed Monte Carlo simulations of the expected backgrounds, considering realistic energy resolutions and thresholds in the detector.  
In  a low-energy window of 2--30\,keV, where the sensitivity to solar pp and $^7$Be-neutrinos is highest, an integrated pp-neutrino rate of 5900 events can be reached in a fiducial mass of 14\,tons of natural xenon, after 5 years of data. The pp-neutrino flux could thus be measured with a statistical uncertainty around 1\%,  reaching  the precision of solar model predictions.  These low-energy solar neutrinos will be the  limiting background to the dark matter search channel for WIMP-nucleon cross sections below $\sim$2$\times$10$^{-48}$\,cm$^2$ and WIMP masses around 50\,GeV$\cdot$c$^{-2}$, for an assumed 99.5\% rejection of electronic recoils due to elastic neutrino-electron scatters. Nuclear recoils from coherent scattering of solar neutrinos will limit the sensitivity to WIMP masses below $\sim$6\,GeV$\cdot$c$^{-2}$ to cross sections above $\sim$4$\times$10$^{-45}$cm$^2$.
DARWIN could reach a competitive half-life sensitivity of 5.6$\times$10$^{26}$\,y  to the neutrinoless double beta decay of $^{136}$Xe  after 5 years of  data, using 6\,tons of natural xenon in the central detector region.
}

\begin{document}
\maketitle
\flushbottom

\section{Introduction}

As several large-scale xenon-based dark matter detectors are under development, the question on their physics capabilities beyond direct dark matter detection arises. Here we consider other rare-events searches, such as the measurement of low-energy solar neutrinos, of coherent neutrino-nucleus scattering and the search for the neutrinoless double beta decay of $^{136}$Xe, by studying the concrete example of the proposed DARWIN (Dark matter WIMP search with noble liquids) \cite{Baudis:2012bc,Schumann:2011ts,darwin}  experiment. Other large liquid xenon detectors aiming to observe solar neutrinos and double beta decays have been discussed in the past \cite{Suzuki:2000ch,Arisaka:2008mb}.

Although it is firmly established that neutrinos change flavour when 
traveling over macroscopic distances, and the oscillation parameters of solar neutrinos are measured with certain precisions, the pp-neutrino flux has never been observed in real time \cite{Antonelli:2012qu,Robertson:2012ib}. The pp neutrinos originate in the main fusion reaction powering our Sun and their predicted flux is much larger than that of neutrinos produced in subsequent reactions. If the solar luminosity constraint is imposed, their flux can be predicted with a 1\% precision \cite{bahc05}.  Using all existing solar and terrestrial neutrino experimental data, but relaxing the solar luminosity constraint, 
the pp-neutrino flux can be determined with a 1-$\sigma$ uncertainty of about 10\% in a solar model independent analysis \cite{GonzalezGarcia:2012sz}. These constraints on the neutrino-inferred luminosity, which agree with the measured value within the same 1-$\sigma$ uncertainty, could be improved if a direct, real-time measurement of the pp-flux becomes available. Additionally, the direct observation of pp neutrinos can probe the vacuum-dominated oscillation probability of sub-MeV neutrinos, yielding, along with the measured $^{7}$Be neutrino flux by Borexino \cite{Bellini:2011rx}, further tests of the MSW-LMA solution of the solar neutrino problem.  We show that DARWIN could reach a precision in the pp flux measurement of $\sim$1\% after 5 years of data.

The observation of the neutrinoless double beta decay would prove that neutrinos are massive Majorana particles and would provide valuable information about the absolute neutrino mass scale and mass hierarchy. Current searches constrain the half-life of the decay to values larger than about $(2-3)\times10^{25}$\,y \cite{Agostini:2013mzu,GomezCadenas:2011it,Giuliani:2012zu,Zuber:2012ij}. In particular, searches using xenon in its liquid form, such as EXO, or mixed with a liquid scintillator, such as KamLAND-Zen, recently published lower limits on the $^{136}$Xe half-life  of $1.6\times10^{25}$\,y (90\% C.L.) \cite{Auger:2012ar} and $1.9\times10^{25}$\,y (90\% C.L.) \cite{Gando:2012zm}, respectively. While improvements are expected with more data, and future phases of these projects, as well as with other experiments using xenon, such as NEXT \cite{MartinAlbo:2013ve}, we show that {\sc DARWIN} could have a non-negligible sensitivity to this physics channel even without isotopic enrichment. 

{\sc DARWIN} is an R\&D and design study for a next-generation dark matter detector using liquid xenon (LXe) and possibly liquid argon (LAr) as  detection media for Weakly Interacting Massive Particles (WIMPs).  Building upon the experience with current and near-future xenon and argon detectors such as XENON100 \cite{Aprile:2011dd}, XENON1T \cite{Aprile:2012zx}, ArDM \cite{Badertscher:2013ygt} and DarkSide \cite{Wright:2011pa}, it will employ large, unsegmented target masses (about  20\,t and 30\,t of LXe and LAr, respectively) capable of fiducialization in dual-phase time projection chambers.  The cryostats containing the noble liquids are immersed in large water Cherenkov shields  to suppress the background originating from environmental radioactivity and muon induced neutrons to negligible levels.  The primary aim of {\sc DARWIN} is to scan the WIMP-nucleon  cross section region down to 10$^{-48}$cm$^2$, and  to yield a high-statistics measurement of WIMP-induced nuclear recoils, should a first signal be seen by a current or near-future dark matter detector.  The goal is to achieve energy thresholds of  2\,keV and 7\,keV in xenon and argon respectively, which translate into roughly  6.6\,keV$_{\rm nr}$ and 30\,keV$_{\rm nr}$ thresholds for observing nuclear recoils as expected from WIMP interactions,  and background levels of $\sim$10$^{-4}$ events/(kg$\cdot$keV$\cdot$d) in the dark matter region of interest. Such a large, homogeneous, ultra-low background and low-threshold detector is not only tailored to search for WIMPs, but opens the possibility to conduct other rare-event searches, some of which we explore in detail in this paper.  

In section \ref{sec:event_rates} we predict the expected event rates from low-energy solar neutrinos. In section \ref{sec:backgrounds} we discuss the background requirements for the solar 
neutrino observation, based on detailed Monte Carlo simulations and derive realistic detection rates. In section \ref{sec:nu_backgrounds} we address the question of the neutrino background for {\sc DARWIN}'s main physics channel, the dark matter search, including coherent neutrino-nucleus scattering, and compare with expectations from other background sources. In Section \ref{sec:bb_signal} we estimate {\sc DARWIN}'s sensitivity to observe the neutrinoless double beta decay of $^{136}$Xe and discuss the background requirements for this  higher energy channel. The final section contains a brief summary and a discussion of our main results.

\section{Expected solar neutrino event rates}
\label{sec:event_rates}

In {\sc DARWIN}, solar neutrinos are detected by the elastic neutrino-electron scattering reaction 
\begin{equation}
\nu +  e^- \rightarrow \nu +  e^- ,
\end{equation}
which proceeds through $Z-$ and $W-$exchange for electron neutrinos, and only through neutral current reactions for neutrinos of other flavours.
The pp neutrinos, produced in the center of our Sun via the proton fusion reaction p + p $\rightarrow$ $^2$H + e$^{+}$ + $\nu_e$,  
 have a continuous energy spectrum with an endpoint at 420\,keV.  The  $^7$Be-neutrinos, born in the electron-capture reaction 
 $^7$Be + e$^{-} \rightarrow $$^7$Li + $\nu_e$, are mono-energetic with energies of 383\,keV (BR=10\%) and  862\,keV (BR=90\%).  The first, subdominant branch, occurs when the reaction proceeds through an excited state of $^7$Li. 
 The maximum electron recoil energies deposited in the target are 261\,keV,  226\,keV and 665\,keV for  pp and the two $^7$Be-neutrino branches, respectively \cite{bahc89}.

We consider here the xenon part of {\sc DARWIN}, for in argon the electronic recoil spectrum at low energies will be dominated by beta-decays from $^{39}$Ar, with a half-life of T$_{1/2}$= 269\,y and an endpoint at 565\,keV. Atmospheric argon has  a ratio of  $^{39}$Ar /$^{nat}$Ar of 8$\times$10$^{-16}$, and  the count rate from $^{39}$Ar in natural argon is typically 1\,kHz/t. While depletion factors of larger than 100 have been achieved by using argon extracted from underground gas wells \cite{Back:2012pg},  this background component will be too high even in the optimistic case when depletion factors above 10$^3$ compared to natural argon are reached.

We note that, unlike a hypothetical WIMP, which produces nuclear recoils (NRs) by WIMP-xenon collisions, solar neutrinos create so-called electronic recoils (ERs) when elastically colliding with electrons in the medium. These are {\sl a priori} indistinguishable from single interactions coming from beta decays or from the gamma background. The solar neutrino signal in {\sc DARWIN} will be dominated by pp neutrinos, followed by  $^7$Be and CNO neutrinos. The latter contribute with a rate which is about a factor of 10 below the one from $^{7}$Be neutrinos, thus we neglect CNO neutrinos here. Solar neutrinos originating from other reactions have much lower fluxes \cite{bahc05,SSM}, and would yield a negligible contribution to the signal below an electronic recoil energy of $\sim$\,600\,keV. 

We calculate the expected rates from pp and $^7$Be neutrinos under the following assumptions:  a total LXe mass of 21.4\,t, an instrumented LXe mass of 18\,t and a fiducial xenon mass of 14\,t of LXe (see Section \ref{sec:backgrounds}), corresponding to  N$_0$=3.47$\times$10$^{30}$ target electrons, and an energy threshold for detecting the recoiling electrons of 2\,keV. Given a scintillation light yield in liquid xenon of $\sim$ 46\,photons/keV, this is a realistic energy threshold considering the light yields already achieved in dual-phase TPCs such as XENON100 \cite{Aprile:2011dd} and LUX \cite{Akerib:2012ys}.  
The mean cross sections for elastic neutrino-electron scattering at these low energies are \cite{bahc89}:

\begin{equation}
\sigma_{\nu_{e}} = 11.6 \times 10^{-46} {\rm cm^2} \hspace*{2mm} {\rm (pp)}, \hspace*{3mm}  \sigma_{\nu_{e}} = 57.9 \times 10^{-46} {\rm cm^2} \hspace*{2mm} {\rm (^7Be)}.
\end{equation}
For the total neutrino fluxes, we consider the lowest and hence more conservative values from \cite{bahc05}, which lists the predicted fluxes for seven distinct solar models:

\begin{equation}
\phi_{pp} = 5.94 \times 10^{10} {\rm cm^{-2} s^{-1}},  \hspace*{3mm} \phi_{^{7}Be} = 4.38 \times 10^{9} {\rm cm^{-2} s^{-1}}.
\end{equation}
To calculate the expected differential electron recoil spectrum, $dN_e(E_e)/dE_e$,  the differential cross section 
$d\sigma(E_e,E_{\nu})/dE_e$ is convoluted with the 
incoming neutrino spectrum $dN_{\nu}(E_{\nu})/dE_{\nu}$.  The neutrino spectral shape can be written as \cite{raffelt_book}:

\begin{equation}
\frac{dN_{\nu}(E_{\nu})}{dE_{\nu}} = A (Q + m_e - E_{\nu}) \left[(Q + m_e -E_{\nu})^2 - m_e^2\right]^{\frac{1}{2}} E_{\nu}^2 \,F,
\end{equation}
where $Q$  is the maximum neutrino energy,   $E_{\nu}$ is the initial neutrino energy,   $m_e$ is the mass of the electron, 
$F$ is a correction factor close to unity for the low-$Z$ nuclei in the Sun, and $A$ is a normalization factor. The differential cross section is \cite{hooft1971}:

\begin{equation}
\frac{d\sigma(E_e, E_{\nu})}{dE_e} = \frac{2 G_F^2 m_e}{\pi}  \left[ g_L^2 + g_R^2\left(1 - \frac{E_e}{E_{\nu}}\right)^2 - g_L g_R \frac{m_e}{E_e E_{\nu}}\right],
\end{equation}
where $E_e$ is the kinetic energy of the final state electron,  $G_F$ is the Fermi constant and 

\begin{equation}
g_L = \pm \frac{1}{2} + \sin^2 \theta_W, \hspace*{3mm} g_R = \sin^2\theta_W,
\end{equation}
where $\theta_W$ is the Weinberg angle with sin$^2\theta_W$ = 0.231, and the sign conventions for $g_L$ is positive for $\nu_e$ and negative for 
$\nu_{\mu,\tau}$. The scattering cross section for $\nu_e$ is roughly six times higher compared to the one for the two other neutrino flavors.

To determine the actual scattering rates in a detector, the survival probability, which is a function of the measured neutrino oscillation parameters and of the neutrino energy, is to be taken into account.  For low-energy  pp- and $^{7}$Be-neutrinos, the LMA-MSW solution of the solar neutrino problem predicts vacuum-dominated oscillations.  The probability for the neutrino to arrive at the detector as an electron neutrino can  be approximated as \cite{bachall2004}:

\begin{equation}
P_{ee} = \cos^4 \theta_{13} \left(1-\frac{1}{2}\sin^2 2\theta_{12} \right)+\sin^4\theta_{13}, 
\end{equation}
where $\theta_{13}$ and $\theta_{12}$ are the mixing angles between the first and third, and the first and second generation of neutrinos, respectively. Using the most recent measured central values for these mixing angles,  
sin$^2\theta_{12}$=0.312 \cite{Fogli:2011qn} and sin$^22\,\theta_{13}$=0.092 \cite{An:2012eh} we obtain an electron-neutrino survival probability of P$_{ee}$=0.543 and the following overall  rates, considering an energy threshold of 2\,keV in our detector:

\begin{equation}
R_{pp} \simeq  1.05\,\,{\rm events}/({\rm t}\cdot {\rm d}), \hspace*{3mm} R_{^7Be} \simeq  0.51\,\,{\rm events}/({\rm t}\cdot {\rm d}). 
\end{equation}

\begin{figure}[tbp]
\includegraphics[scale=0.38]{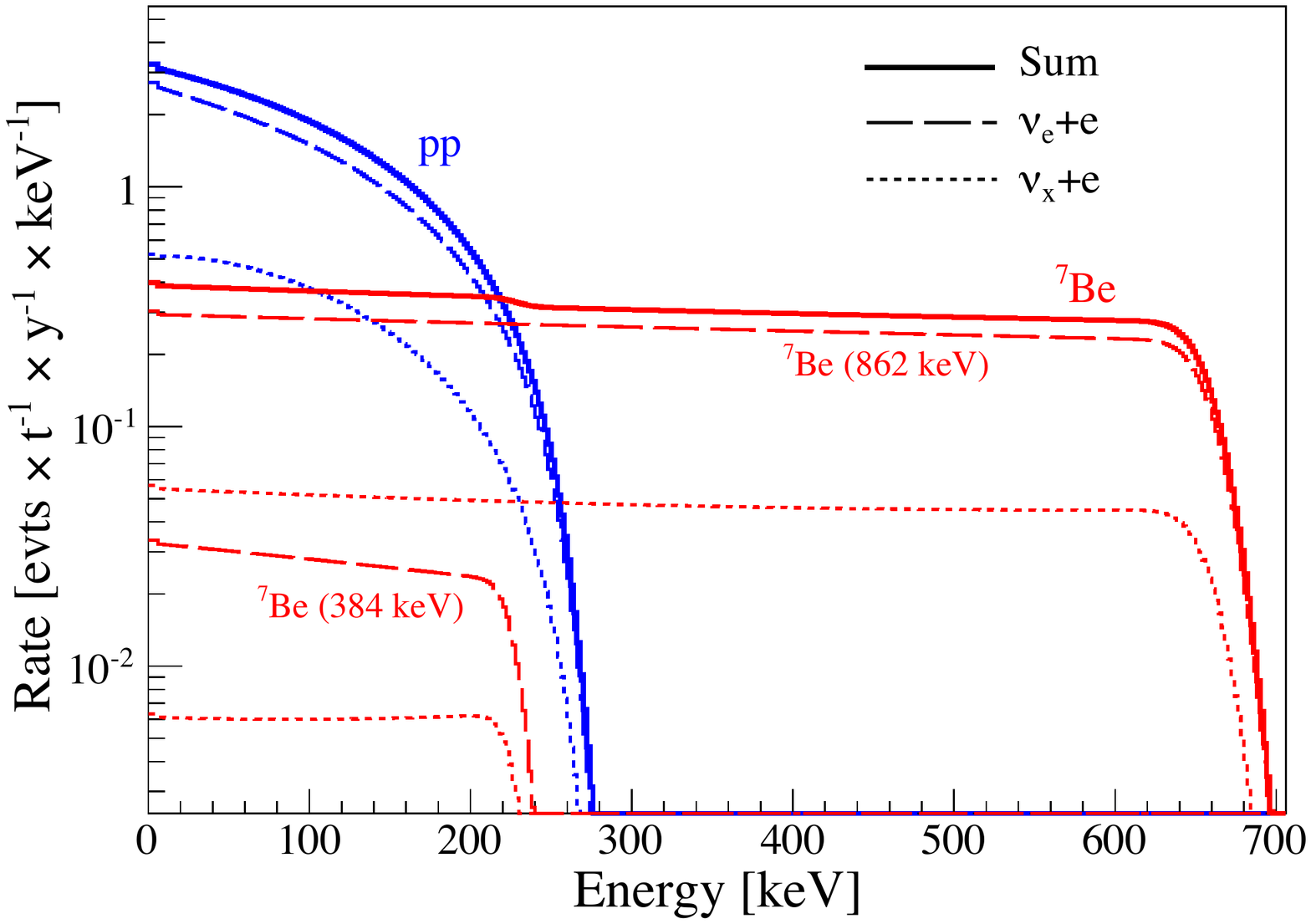}
\includegraphics[scale=0.38]{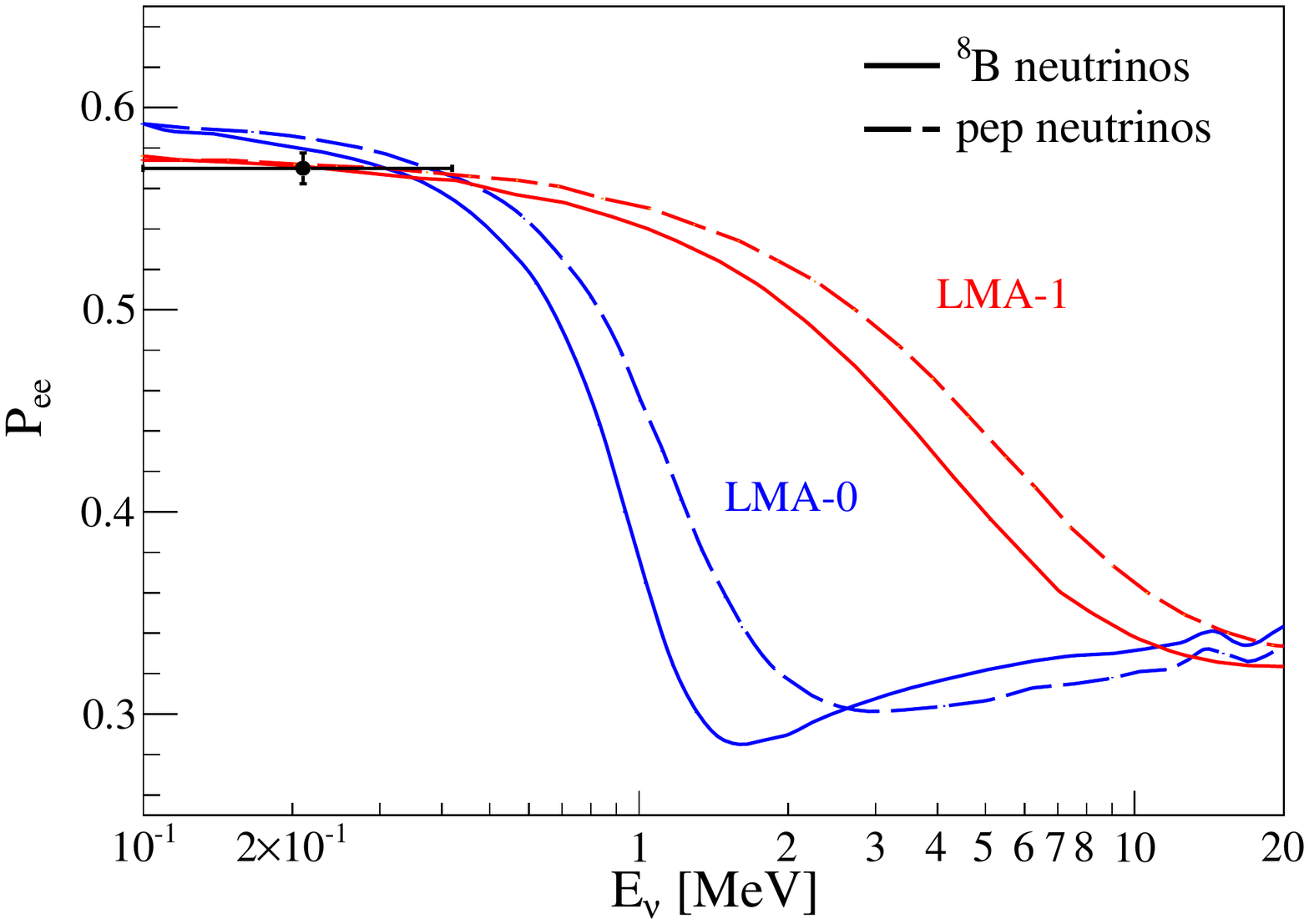}
\caption{\small{(Left): Differential electron recoil spectra in liquid xenon for pp- (blue) and $^7$Be (red) neutrinos. The dashed lines show the contribution from electron neutrinos, the dotted lines the ones from other flavours, while the solid lines represent their sum. (Right): Survival probability of electron neutrinos as a function of energy  for pep (dashed) and $^8$B (solid) for two different oscillation scenarios LMA-0 (blue) and LMA-I (red), where the relative difference is expected to be of similar size for pp-neutrino  \cite{Friedland:2004pp}. The datapoint shows the result of a possible pp-neutrino measurement with {\sc DARWIN}, in a fiducial LXe mass of 14\,t. The horizontal error bar represents the energy spread for the pp-neutrinos and the vertical error bar shows the expected statistical error for 5900 events.}}
\label{fig:survival}
\end{figure}

The differential electron recoil spectrum, $dN_e(E_e)/dE_e$ gives the number of neutrino-electron elastic scattering events within the electron recoil kinetic energy 
 interval $E_e$ to $E_e$+$dE_e$. Due to neutrino flavour conversion, more than one neutrino flavour may arrive at the detector. Nonetheless, the final state scattered 
 electrons will be experimentally indistinguishable from one another and the contributions of the various neutrino flavours must be taken into account in the calculation of the differential energy spectrum:
 
 \begin{equation}
 \frac{dN_e(E_e)}{dE_e} = N_0 \times  t  \times \sum_{i}^{flavours} \int dE_{\nu} \frac{d\phi_i(E_{\nu})}{dE_{\nu}} \frac{d\sigma_i(E_e, E_{\nu})}{dE_{e}},
 \end{equation}
where $N_0$ is the number of electrons in the target, $t$ is the measuring time and the sum goes over all active neutrino flavours.  Figure\,\ref{fig:survival}, left,  shows the differential electron recoil spectra in the detector,  assuming an energy resolution as achieved by the XENON100  experiment \cite{Aprile:2011dd}.

The real-time measurement of pp-neutrinos  will test the main energy production mechanism in the Sun, for the pp- and $^7$Be-neutrinos together account for more than 98\% of the total neutrino flux predicted by the Standard Solar Model. The flux of $^7$Be-neutrinos has been measured precisely by the Borexino experiment to be $(2.78\pm0.13)\times10^9$\,cm$^{-2}$s$^{-1}$\,\cite{Bellini:2011rx} assuming pure $\nu_e$ neutrinos. However, the most robust predictions of the Standard Solar Model\,\cite{SSM} are for the pp-neutrino flux, which is heavily constrained by the solar luminosity. Assuming 14\,t  LXe fiducial mass and an energy region of interest 2--30\,keV,  about  5900  pp-neutrinos are expected in DARWIN after 5 years of data taking.  The upper energy boundary is motivated by the energy at which the solar neutrino induced electron recoil spectrum and the two-electron spectrum from the $2\nu\beta\beta$-decay of $^{136}$Xe intersect, as shown  in Figure \ref{fig:diffspectra}, left.  
{\tr{The systematic error on this measurement, from the error in the determination of the fiducial volume with a LXe mass of 14\,t, was estimated by a Monte Carlo simulation of neutrino scatters uniformly distributed in the active liquid xenon target. Assuming a radial position resolution of $\sigma$=15\,mm, which is a realistic goal based on existing measurements \footnote{Resolutions of about 3\,mm and 5\,mm were reached in the XENON100 and LUX TPCs, with 1-inch and 2-inch light sensors, respectively.} and Monte Carlo predictions with 3-inch light sensors \footnote{Monte Carlo simulations of the expected position resolution with 3-inch sensors and a gas-gap of 6\,cm predict a 1-sigma x-y position uncertainty of 8\,mm. The z-position uncertainty, which in a TPC is determined from the time difference between the prompt, and delayed scintillation light signals is well below 1\,mm.}, the error in the fiducial volume calculation is $<$0.1\%, which is well below the statistical uncertainty of the measurement. The systematic uncertainty in the determination of the event vertex will be tested by using collimated external sources, such as $^{57}$Co \cite{Aprile:2011dd}, internal, point-like alpha sources, as well as short-lived radioactive isotopes mixed into the LXe, such as $^{83m}$Kr \cite{Manalaysay:2009yq}. The latter leads to a constant volume activity throughout the active volume and deviations can be attributed to systematic biases in the vertex determination.}}

Such a large statistics measurement of the pp-neutrino flux will open the possibility to distinguish among different oscillation scenarios. For example, non-standard neutrino interactions  can modify the survival probability of electron neutrinos in the vacuum to matter dominated oscillation transition region around 1\,MeV but also at lower, pp-neutrino energies. Figure\,\ref{fig:survival}, right, shows the survival probability of electron neutrinos as function of energy  for pep and $^8$B for two different oscillation scenarios, LMA-0 and LMA-I, the relative difference for pp-neutrinos being of similar size\,\cite{Friedland:2004pp}. The datapoint represents a possible measurement of the survival probability of pp-neutrinos in {\sc DARWIN}, with a relative statistical error of 1.3\%. The total neutrino rate per year, taking into account the 2\,keV energy threshold for detecting the recoiling electrons and an upper energy threshold of 30\,keV is 1331 events (see also Table \ref{tab:rates}).

\section{Background considerations}
\label{sec:backgrounds}

Elastic neutrino-electron interactions will cause electronic recoils competing with a potentially much larger background coming from detector construction materials, from radioactive noble gas impurities in  the liquid xenon itself, such as  $^{85}$Kr and $^{222}$Rn, and from double beta decays of $^{136}$Xe. With an instrumented water Cherenkov shield, ultra-low radioactivity construction materials,  the self-shielding of liquid xenon and the information on the spatial event-position, {\sc DARWIN} will be designed to reduce most of this background noise to extremely low levels.  We have modeled the known background contributions using the Geant4 \cite{geant4} Monte Carlo simulation package and a simplified detector geometry.  It consists of a double-walled copper cryostat, a cylindrical teflon TPC structure, two photosensor arrays, one in the liquid below the target and one in the vapour phase above the liquid, a  diving bell to maintain a constant LXe level, as well as five electrodes made of copper, defining the potentials in the TPC. The 2.8-inch diameter photosensors are modeled after the {\sc QUPID} \cite{Teymourian:2011rs} a hybrid avalanche photodiode (APD), and consist of a cylindrical quartz body, a semi-spherical quartz window, a base plate and a central, cylindrical quartz pillar holding the APD.  However, other light and charge readout detectors are being considered for {\sc DARWIN} as well, one photosensor candidate being the new 3-inch Hamamatsu R11410 photomultiplier tube, to be used in XENON1T and extensively tested in liquid xenon \cite{Baudis:2013xva}. Table \ref{tab:materials} summarizes the materials, masses and radioactivity levels of the detector components used in the simulation. The specific activities are typical numbers from existing measurements \cite{Baudis:2011am,Aprile:2011ru}.

\begin{table}[h!]
\centering
\begin{tabular}{|llrr|}
\hline
Detector component      & Material    & Mass [unit]   &    Specific activity [mBq/unit]\\
&& &$^{228}$Th/$^{226}$Ra/$^{40}$K/$^{60}$Co\\
\hline
Outer cryostat         & Copper  & 2150\,kg  &    0.02/0.07/0.023/0.008 \\
Inner cryostat         & Copper  & 1737\,kg   &    0.02/0.07/0.023/0.008 \\
Electrodes, bell        & Copper  & 113\,kg     &   0.02/0.07/0.023/0.008 \\
TPC					& Teflon    & 313.6\,kg  &    0.2/0.05/0.59/0.01\\
Photosensors			& Quartz   & 1050 pieces &    0.28/3.2/0.39/6.0/0.15\\
\hline
Total (instrumented) volume         & LXe       & 21.4 (18)\,t	&    0.1\,ppt $^{nat}$Kr, 0.1\,$\mu$Bq/kg $^{222}$Rn\\
\hline
\end{tabular}
\caption{\small Main detector components in the mass model of the Monte Carlo simulation of  expected backgrounds from the intrinsic radioactivity of materials. We assume a $^{85}$Kr/$^{nat}$Kr ratio of 2$\times$$10^{-11}$\,mol/mol.} 
\label{tab:materials}
\end{table}

\begin{figure}[!h]
\mbox{
\includegraphics[scale=0.38]{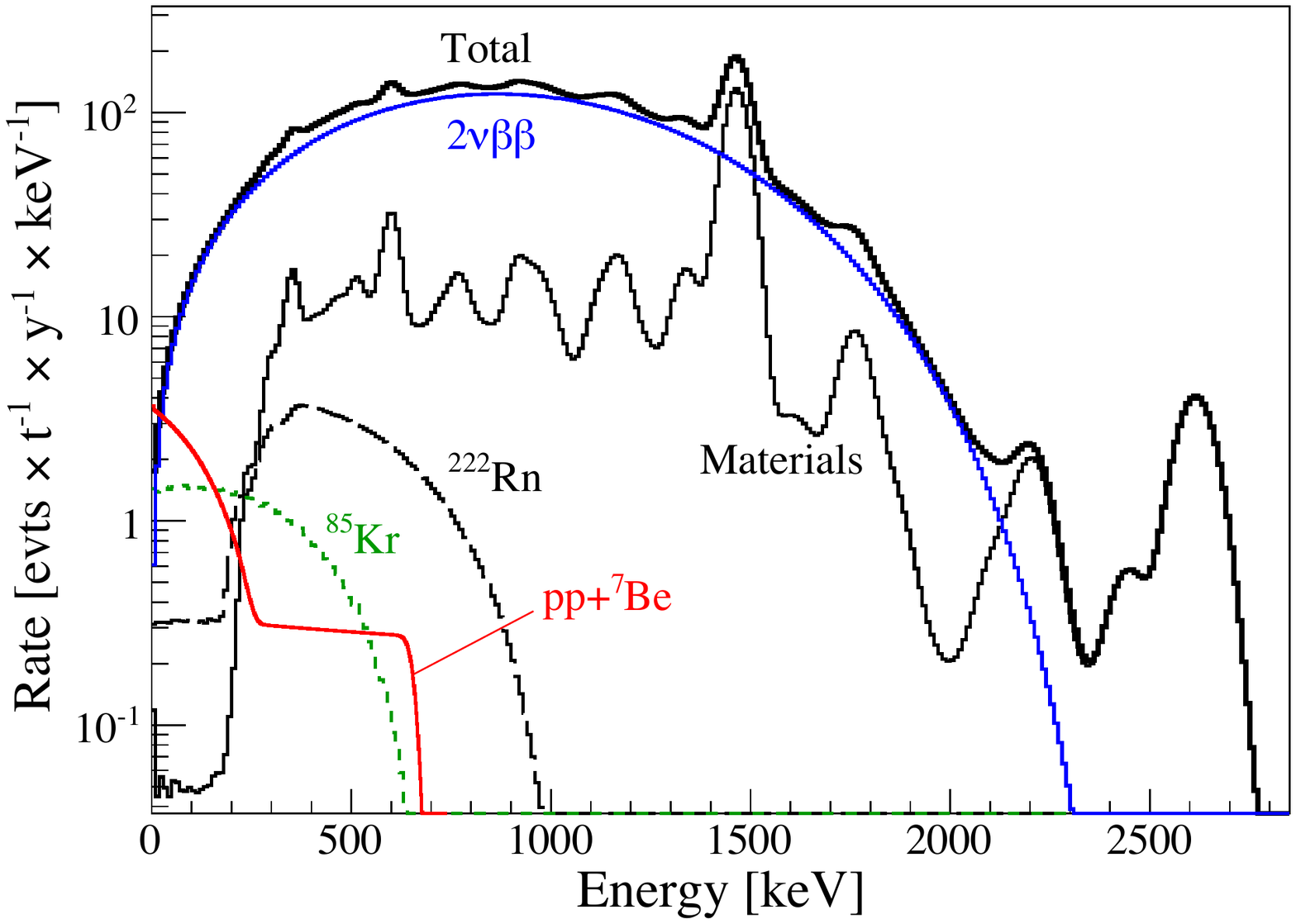}
\includegraphics[scale=0.38]{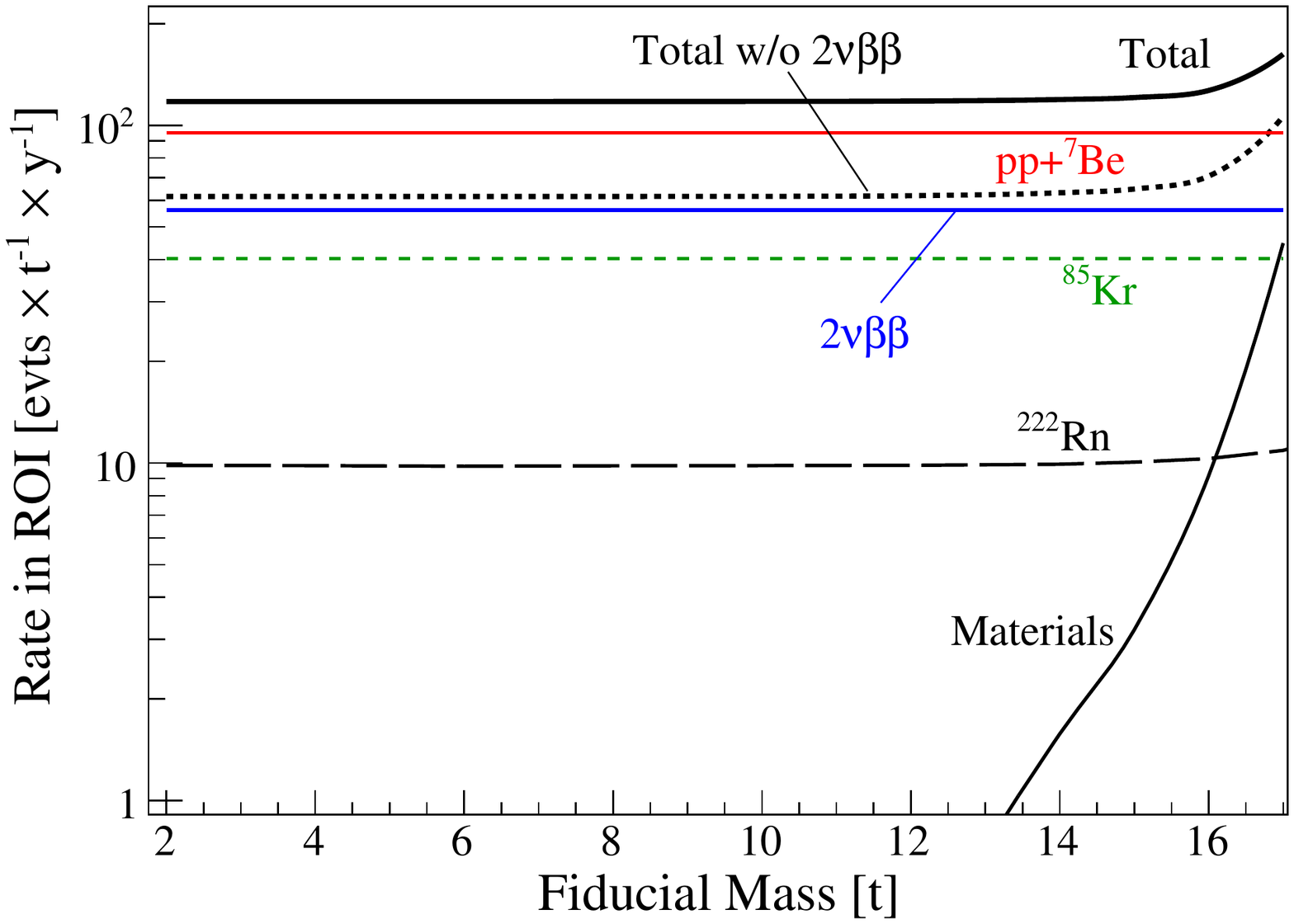}
}
\caption{\small{(Left): Overall predicted background spectrum from detector construction materials (see Table \ref{tab:materials}) and internal ($^{85}$Kr, $^{222}$Rn, $^{136}$Xe) contaminations for a central, 14\,t region of the detector.  The expected background from $^{85}$Kr decays  (green, 0.1\,ppt of natural krypton), from $^{222}$Rn decays  (black dashed,  0.1\,$\mu$Bq/kg) and from $^{136}$Xe 2$\nu\beta\beta$-decays (blue) is also shown separately, along with the total neutrino signal (red, pp and $^7$Be).  (Right): Predicted signal and background rates in the 2-30\,keV energy region as a function of fiducial liquid xenon mass.}}
\label{fig:bg_vs_mass}
\end{figure}

In Figure \ref{fig:bg_vs_mass}, left, we show the overall background spectrum coming from detector construction materials and from radioactivity intrinsic to the liquid xenon, including the  2$\nu\beta\beta$-decay spectrum of $^{136}$Xe, with T$_{1/2}$ = 2.11$\times$10$^{21}$\,y \cite{Ackerman:2011gz}. The non-intrinsic background is dominated by the photosensors, followed by the cryostat, the TPC and the diving bell. In the energy range of 2--30\,keV with no electronic recoil rejection,   {\sc DARWIN} would observe 19\,events/y in a central detector region with fiducial mass of 14\,t.  Assuming 99.5\% rejection of electronic recoils in the energy region 2--10\,keV, relevant for the dark matter search, the contribution from this component is about 3$\times$10$^{-2}$ events/y.

The intrinsic background to xenon as a detection medium poses strong requirements on internal radio-activity levels: a contamination of the liquid xenon with natural krypton of about 0.1\,ppt and a radon level in the liquid of about 0.1\,$\mu$Bq/kg is to be ensured.  This can be achieved by purifying the noble gas with krypton-distillation columns and with ultra-clean, charcoal-based radon filters, and by the use of materials with low radon emanation. For krypton, currently achieved $^{nat}$Kr-levels are (1.0$\pm$0.2)\,ppt in XENON100 \cite{Lindemann:2013kna} (with a detection limit by gas chromatography and mass spectrometry of 0.008\,ppt), (3.5$\pm$1.0)\,ppt in LUX \cite{Akerib:2013tjd} and $<$2.7\,ppt in the XMASS detector \cite{Abe:2013tc}. In the case of radon, the further reduction will be more challenging, albeit not impossible, as $^{222}$Rn levels of (3.7$\pm$0.4)\,$\mu$Bq/kg were measured in EXO-200 \cite{Albert:2013gpz}, and levels around 20\,$\mu$Bq/kg and 10\,$\mu$Bq/kg were achieved in XENON100 \cite{Aprile:2011vb}, LUX \cite{Akerib:2013tjd}, and XMASS \cite{Abe:2013tc}, respectively. Assuming a 99.8\% rejection of $^{214}$Bi events using the $^{214}$Bi-$^{214}$Po coincidence (see Section \ref{sec:bb_signal}), the radon-induced background in this low-energy region can be reduced by about 30\%.

The contribution of these internal background components is shown separately in Figure~\ref{fig:bg_vs_mass}: they yield a total rate of $\sim$700\,events/y in the energy region 2--30\,keV and a rate of about 1\,event/y in the energy region 2--10\,keV, assuming 99.5\% rejection of electronic recoils in the latter case. This rate is comparable to the background coming from $^{136}$Xe 2$\nu\beta\beta$-decays.
In Figure~\ref{fig:bg_vs_mass}, right, we show the background and signal rates as a function of fiducial liquid xenon mass, motivating our conservative choice of 14\,t of LXe in the central detector region. 

Table~\ref{tab:rates} gives an overview of the most relevant background contributions to the overall event budget. The total background in the energy region 2--30\,keV is dominated by 2$\nu\beta\beta$-decays of $^{136}$Xe, followed by decays of $^{85}$Kr and $^{222}$Rn. While the latter can be in principle further reduced by noble gas purification,  the solar neutrino measurement requires the subtraction of the 2$\nu\beta\beta$-decay component. To diminish its contribution without the need of background subtraction, and to extend the energy range over which the solar neutrinos can be observed beyond the 30\,keV upper bound, one might consider using xenon gas that is depleted in the $^{136}$Xe isotope.  The overall background without the 2$\nu\beta\beta$ component is shown in the same figure.

\begin{table}[tbp]
\centering
\begin{tabular}{|lcc|}
\hline
Physics channel                 &  Low-energy $\nu$ measurement  & Dark matter search   \\
                       %  \hline
                        
Energy range & 2--30\,keV                    & 2--10\,keV \\
Assumptions & No ER/NR discrimination & 99.5\% ER rejection \\
                         &                                                &50\% NR acceptance\\
 Rate                         &  Events/(14\,t$\cdot$y)  & Events/(14\,t$\cdot$y)   \\
 \hline
Solar pp neutrinos & 1180 &  1.75 \\
Solar $^{7}$Be neutrinos & 151 &  0.25 \\
%\hline
100\,GeV/c$^2$ WIMP, 2$\times$10$^{-47}$cm$^2$ & 42 & 19\\
40\,GeV/c$^2$ WIMP,   2$\times$10$^{-48}$cm$^2$ & 5.2  & 2.5\\
%\hline
Coherent $\nu$ scattering & 0.98 & 0.45\\
Detector components & 19 & 0.03\\
$^{85}$Kr in LXe (0.1\,ppt of $^{nat}$Kr)& 565 & 0.82\\
$^{222}$Rn in LXe (0.1\,$\mu$Bq/kg) & 139 & 0.20\\
$^{136}$Xe ($2\nu\beta\beta$)& 785 & 0.41 \\
\hline
\end{tabular}
\caption{\small Total solar neutrino and WIMP signal rates in {\sc DARWIN} for a fiducial mass of 14\,t of LXe and two WIMP masses and cross sections.  The energy regions 2-30\,keV and  2-10\,keV are considered for the solar neutrino and the dark matter search, respectively. For the latter, a 99.5\% rejection of electronic recoils, and a 50\% acceptance of nuclear recoils is assumed.}
\label{tab:rates}
\end{table}

\section{Neutrino backgrounds for the dark matter search}
\label{sec:nu_backgrounds}

In this section we contemplate the main physics channel of {\sc DARWIN}, the dark matter search, and calculate the backgrounds from elastic neutrino-electron scatters and from coherent neutrino-nucleus interactions.  The expected dark matter signal events are nuclear recoils from elastic WIMP-nuclei collisions.  The charge-to-light ratio, measured independently for each event, is used to suppress 99.5\% of the electronic recoil background. Such a discrimination level was reached by current-generation detectors and could in principle be improved by using higher drift fields, through analysis techniques, or by reducing the acceptance of nuclear recoils.   Table \ref{tab:rates} gives the expected event rates from WIMP interactions for two cases, assuming the standard halo model: an isothermal halo with a local dark matter density of 0.3\,GeV/cm$^3$, a circular velocity of 220\,km/s and an escape velocity of 544\,km/s \cite{rave,Green:2011bv}.  An in-depth study of the scientific dark matter reach of DARWIN was performed in Ref.~\cite{Newstead:2013pea}.

To calculate the expected differential recoil spectra in the detector, we employ the so-called electron-equivalent energy scale, in general determined with gamma- and electron emission from calibration sources \cite{Baudis:2013cca}. For nuclear recoils as generated by WIMP collisions, or coherent neutrino-nucleus interactions, the scintillation and charge signals are suppressed compared to electronic recoils of the same energy. Our energy scale is based on the detection of the primary scintillation signal, as custom in current xenon dark matter experiments. Using the average relative scintillation efficiency, $\mathcal{L}_{\mathrm{eff}}$, of low-energy nuclear recoils in liquid xenon \cite{Aprile:2010um}, an energy threshold of 2\,keV corresponds to a nuclear recoil energy of 6.6\,keV$_{\rm nr}$, while an upper bound of 10\,keV corresponds to 36.8\,keV$_{\rm nr}$.  The energy resolution of the detector is considered as follows: the nuclear recoil spectra are converted into the spectra of observed number of photoelectrons using  
$\mathcal{L}_{\mathrm{eff}}$ and assuming the same light yield at 122\,keV as achieved in the XENON100 detector \cite{Aprile:2012nq}. In a next step, Poisson fluctuations in the number of photoelectrons are applied and the spectra are converted to the electronic recoil energy scale using the measured light yield of low-energy electronic recoils at a drift field of $\sim$0.5\,kV/cm \cite{Baudis:2013cca}, which agrees well with the NEST prediction \cite{Szydagis:2013sih}. We note however that in the future, both light and charge signals will likely be employed to determine the energy scale of electronic and nuclear recoils in a liquefied noble gas detector.

\begin{figure}[!h]
\includegraphics[scale=0.38]{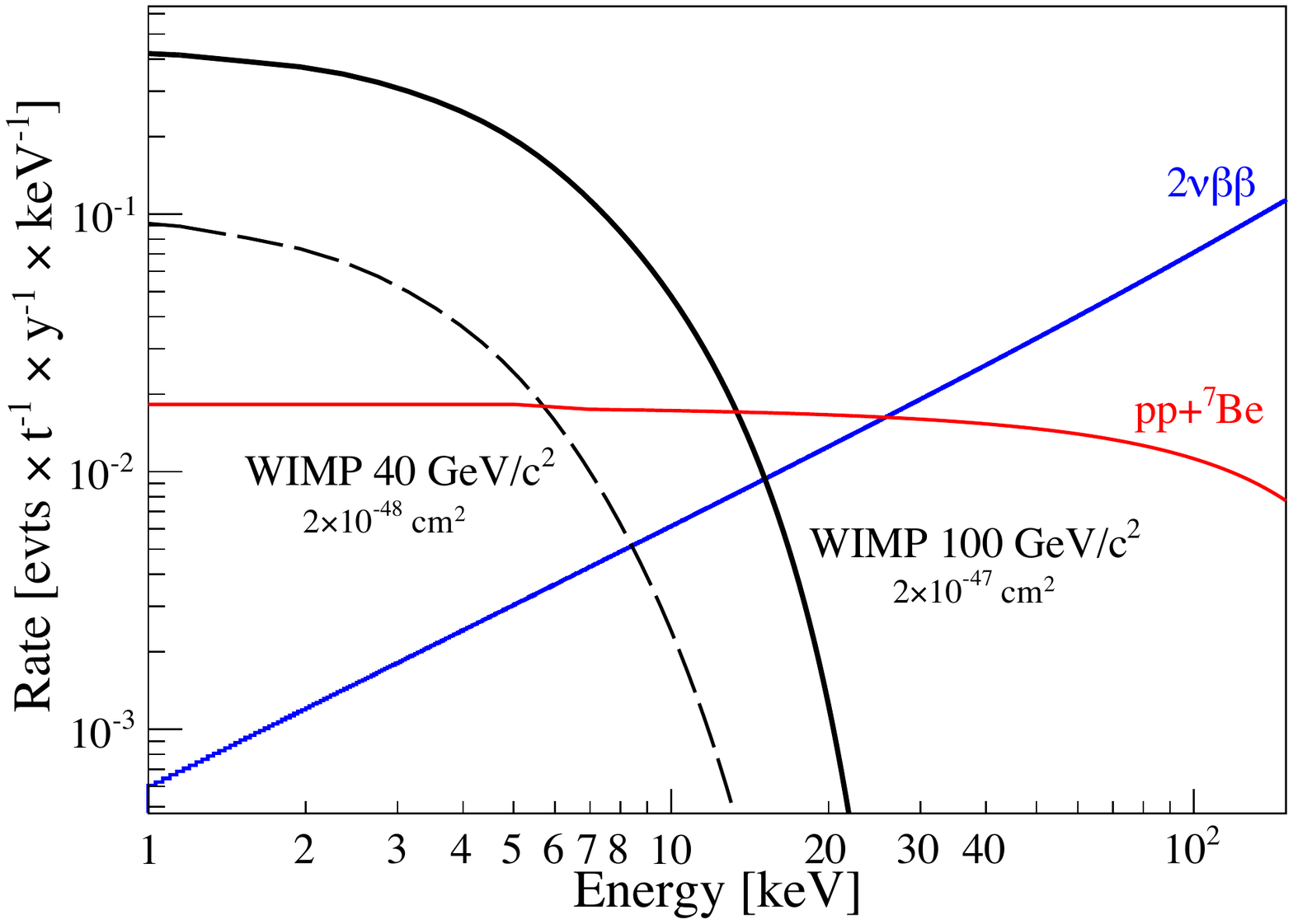}
\includegraphics[scale=0.38]{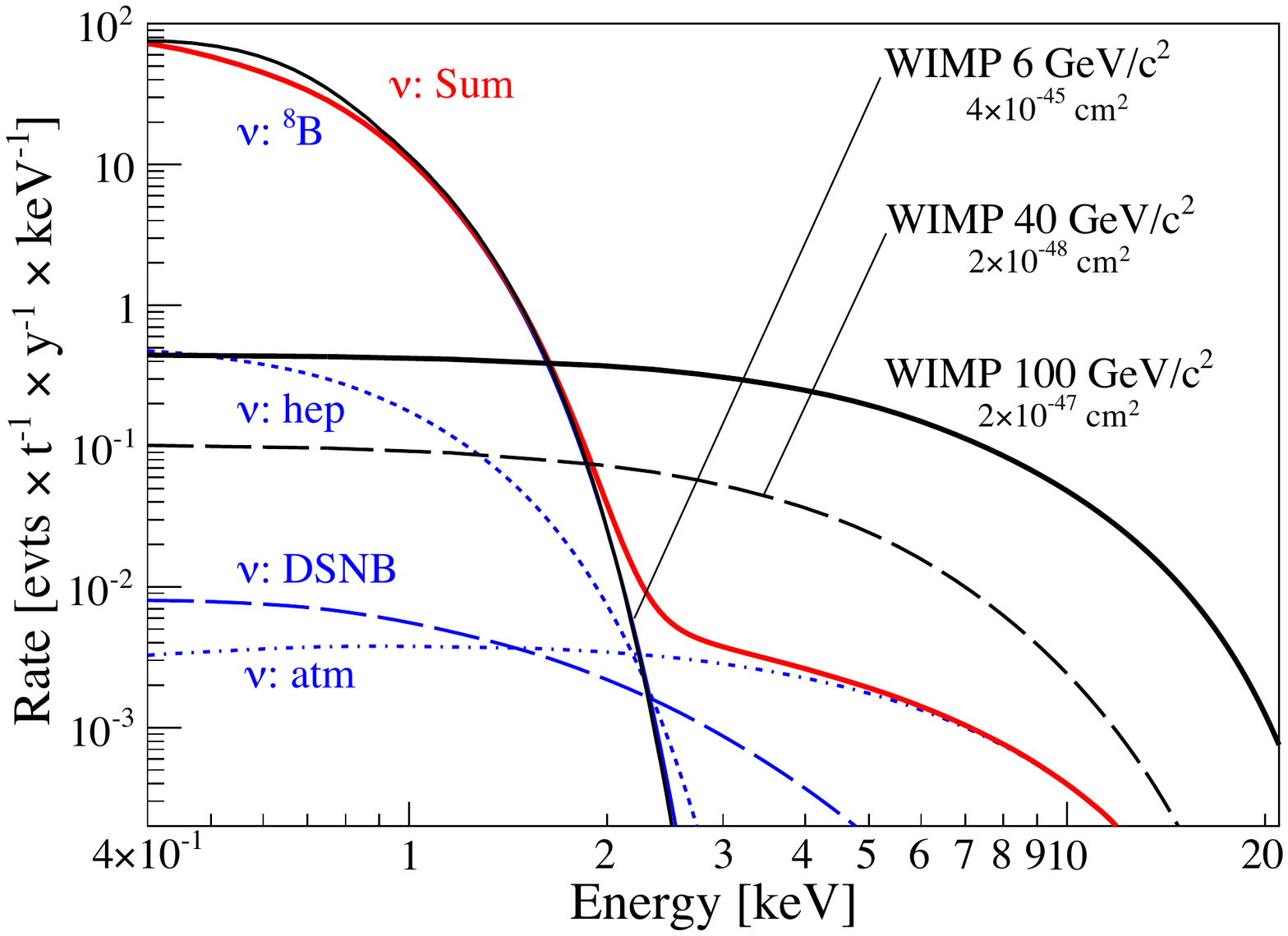}
\caption{\small{(Left): Expected nuclear recoil spectrum from WIMP scatters in LXe for a spin-independent WIMP-nucleon cross section of 2$\times$10$^{-47}$\,cm$^2$ (solid black) and 2$\times$10$^{-48}$\,cm$^2$ (dashed black) and a WIMP mass of 100\,GeV/c$^2$ and 40\,GeV/c$^2$, respectively, along with the summed differential energy spectrum  for pp and  $^{7}$Be neutrinos (red), and the electron recoil spectrum from the double beta decay of $^{136}$Xe (blue), assuming the natural abundance of 8.9\% and the recently measured half life of  2.1$\times$10$^{21}$\,y \cite{Ackerman:2011gz}.  Other assumptions are: 99.5\% discrimination of electronic recoils, 50\% acceptance of nuclear recoils. } (Right): Comparison of the differential nuclear recoil spectrum for various WIMP masses and cross sections (black) to coherent scattering events of neutrinos (red) from the Sun, diffuse supernova background (DSNB), and the atmosphere (ATM). 
The coherent scattering rate will provide an irreducible background for low-mass WIMPs, limiting the cross 
section sensitivity to $\sim 4\times10^{-45}$\,cm$^2$ for WIMPs of 6\,GeV/$c^2$ mass, while WIMP masses 
above $\sim10$\,GeV/$c^2$ will be significantly less affected. For both plots, the nuclear recoil signals are converted 
to an electronic recoil scale and a nuclear recoil acceptance of 50\% is assumed.}
\label{fig:diffspectra}
\end{figure}

In Figure \ref{fig:diffspectra}, left,  we show the expected nuclear recoil energy spectra versus energy for two WIMP masses and cross sections, along with the energy spectra of the recoiling electrons in liquid liquid xenon for the  pp and  $^{7}$Be neutrinos and the double beta spectrum from $^{136}$Xe decays. Assuming a factor 200 discrimination between electronic- and nuclear recoils, and without subtracting the neutrino electronic recoil background, this component becomes a limitation for the dark matter search channel around spin-independent WIMP-nucleon cross sections of 2$\times$10$^{-48}$cm$^2$, dominated by interactions of pp-neutrinos.

The expected nuclear recoil rates from coherent neutrino-nucleus interactions were calculated in \cite{strigari2009,Billard:2013qya,Anderson:2011bi}, for cosmic and terrestrial neutrino sources, respectively. Here we also take into account a realistic energy resolution of the detector, as detailed above.  Coherent neutrino-nucleus elastic scattering has not been observed so far, but the cross section is well predicted in the Standard Model. The largest impact comes from $^8$B and  hep solar neutrinos, as shown in Figure \ref{fig:diffspectra}, right. This figure shows the recoil spectra induced by solar ($^{8}$B and {\sl hep}), atmospheric, and diffuse supernovae neutrinos separately, as well as their sum spectrum. The maximum energy of $^8$B  neutrinos produces nuclear recoils not larger than $\sim$\,4\,keV$_{\rm nr}$ (1.2\,keV), and therefore affects predominantly the sensitivity to low-mass WIMPs. The recoil spectra from atmospheric and diffuse supernovae neutrinos extend to much higher recoil energies, however these neutrinos become relevant at WIMP-nucleon cross sections below 10$^{-48}$cm$^2$  \cite{strigari2009}. A nuclear recoil energy threshold of 6.6\,keV$_{\rm nr}$ (2\,keV), as demonstrated by XENON100 \cite{Aprile:2012nq}, would yield an integrated rate of 0.03\,events/(t$\cdot$y), dominated by atmospheric neutrinos.  A threshold of 1.5\,keV$_{\rm nr}$ (5 electrons), using an energy scale based on the charge signal alone, as demonstrated by XENON10 \cite{Angle:2011th}, would result in 38\,events/(t$\cdot$y). 

\section{Sensitivity to the double beta decay channel}
\label{sec:bb_signal}

\begin{figure}[!h]
\includegraphics[scale=0.38]{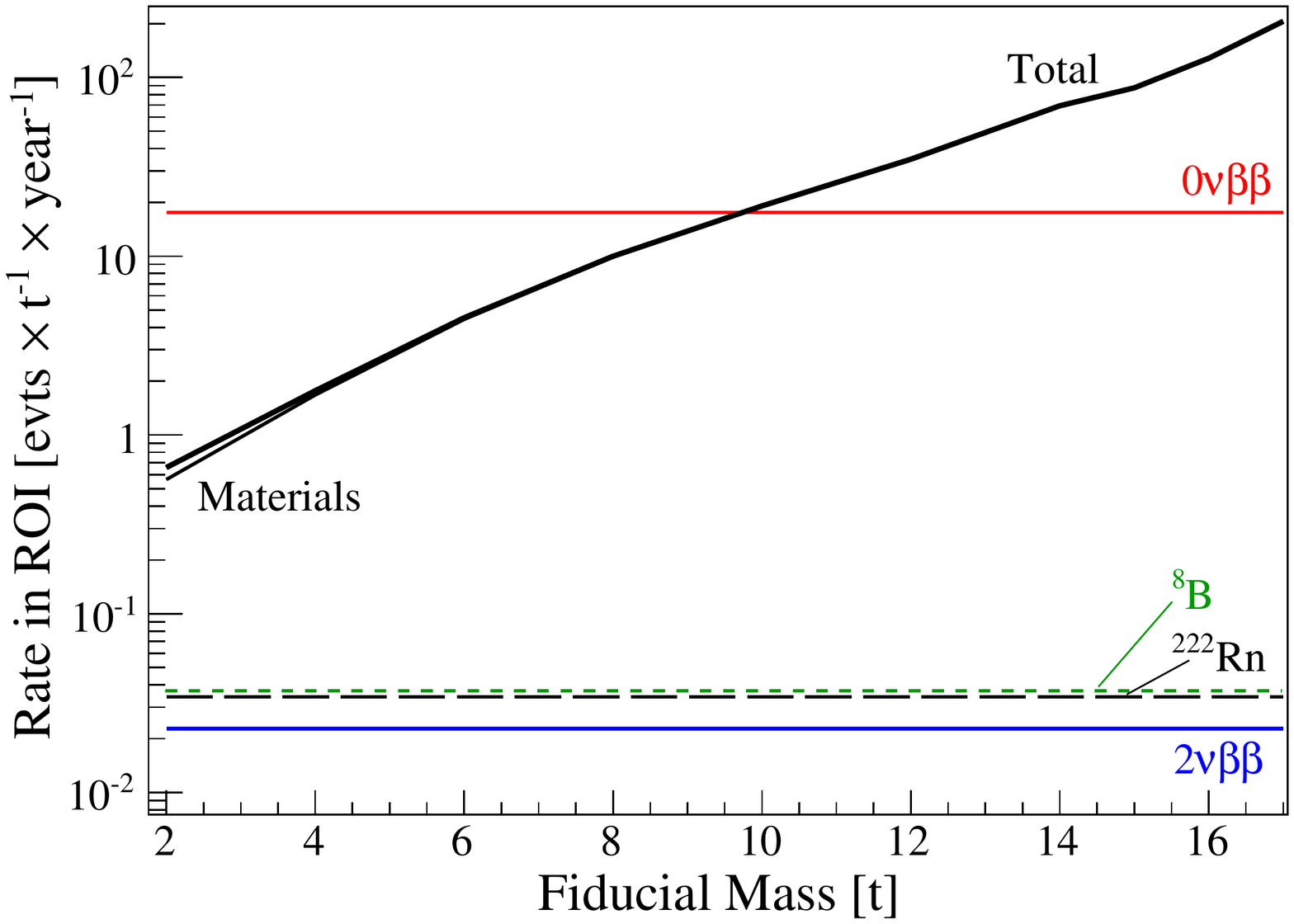}
\includegraphics[scale=0.38]{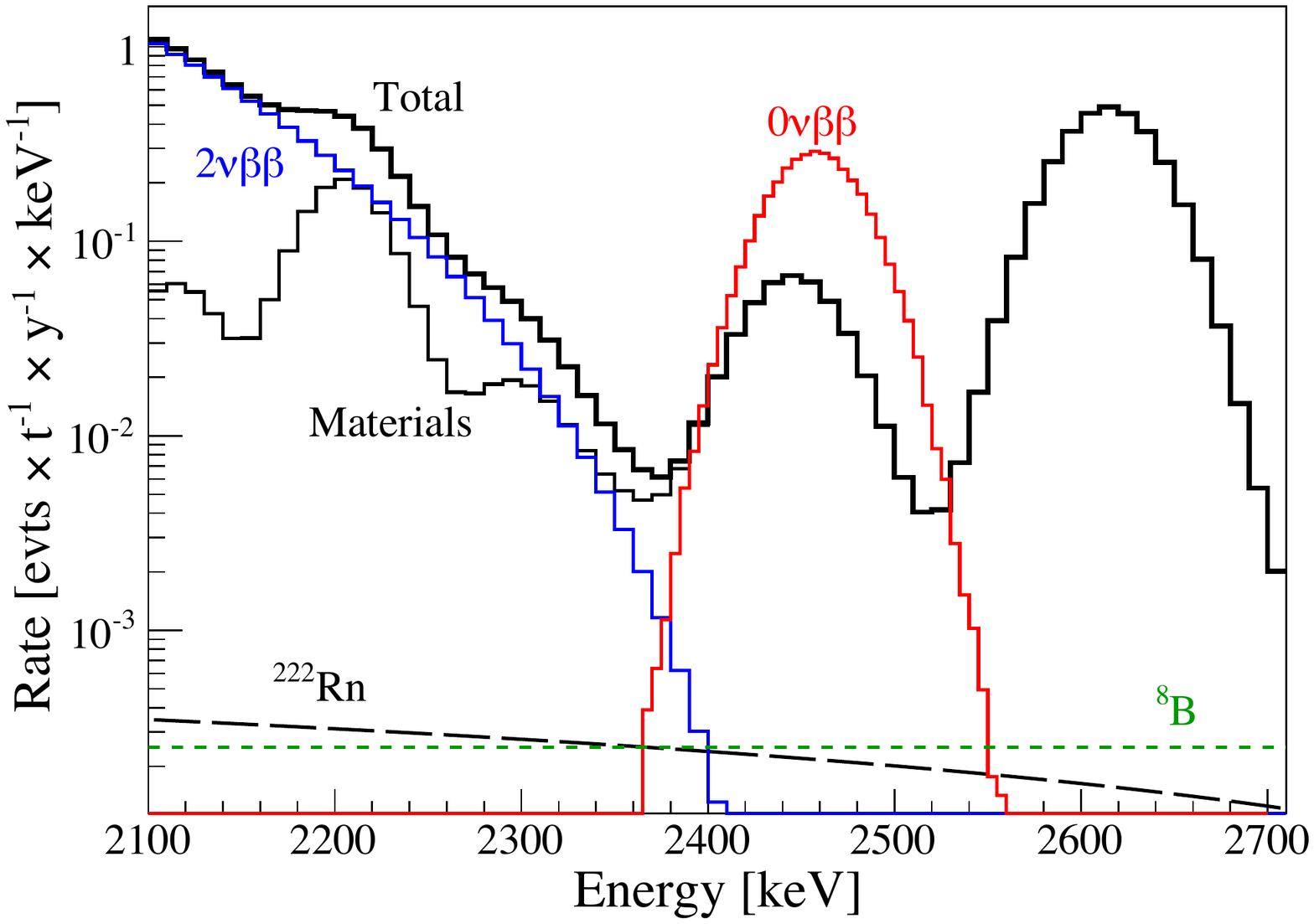}
\caption{\small{(Left): Integral background rate in $\pm$3$\sigma$ energy region around the $Q-$value (2385--2533\,keV) as 
a function of fiducial LXe mass. (Right): Predicted background spectrum around neutrinoless double beta decay peak for 6\,t fiducial mass. We show the overall background (thick  black solid) which includes contributions from detector materials (black), from 0.1\,$\mu$Bq/kg 
of $^{222}$Rn in the LXe (black dashed), from $^{8}$B neutrino scatters (green dotted)and $2\nu\beta\beta$-decays with $T_{1/2}$=2.11$\times$10$^{21}$\,y 
inside the liquid xenon (blue). The potential signal for the neutrinoless double beta decay 
($0\nu\beta\beta$, red) assumes $T_{1/2}$=1.6$\times$10$^{25}$\,y.}} 
\label{fig:doublebeta}
\end{figure}

In the high energy region relevant for the double beta decay search, the fiducial volume cut is less effective in reducing material backgrounds, compared with the low-energy region, as shown in Figure \ref{fig:doublebeta}, left. This is simply due to the longer mean free path of high-energy gammas in liquid xenon. As a compromise between a high $\beta\beta$ source strength (large LXe mass), and a low background, we consider 6\,t of fiducial LXe mass for this physics channel. It contains 534\,kg of $^{136}$Xe for a natural abundance of 8.9\%, thus no isotopic enrichment. 

The $Q-$value of the double beta decay of $^{136}$Xe is (2458.7$\pm$0.6)\,keV \cite{McCowan:2010zz}. Employing an energy scale based on a linear combination of the charge and light signals, which have been shown to be anti-correlated in liquid xenon TPCs \cite{Auger:2012ar,Aprile:2011dd,Manalaysay:2009yq}, the extrapolated energy resolution is $\sigma/E =1\%$ in this high-energy region.  

 {\tr {The combined efficiency of the fiducial volume and multi-scatter cut, which rejects events with a separation larger than 3\,mm in the z-coordinate \cite{Aprile:2011dd} is 99.5\% in a  $\pm$3$\sigma$ energy intervall around the $Q-$value. The materials background is dominated by $^{214}$Bi, followed by $^{208}$Tl decays in the photosensors and in the cryostat, and can only be further reduced, for a given fiducial volume,  by identifying detector construction materials with lower $^{226}$Ra  and $^{228}$Th levels.}} The background contribution from internal radon can be efficiently rejected by so-called $^{214}$Bi-$^{214}$Po tagging. It exploits the fact that the $^{214}$Bi $\beta$-decay (Q$_{\beta}$= 3.3\,MeV) and the $^{214}$Po $\alpha$-decay (Q$_{\alpha}$= 7.8\,MeV) occur close in time, with a mean lifetime of the $^{214}$Po decay of 237\,$\mu$s. At the high energies relevant for the double beta decay, only the $\beta$-decay will contribute to the background. We assume a tagging efficiency of 99.8\%, as achieved in EXO-200 \cite{Albert:2013gpz} and confirmed by us in a Monte Carlo simulation, assuming that $^{214}$Po decays can be detected up to 1\,ms after the initial $^{214}$Bi decay. The event rate from radon, considering the same 0.1\,$\mu$Bq/kg  contamination level as for the dark matter search region, is 0.035\,events/(t$\cdot$y) in a 3$\sigma$ energy region around Q$_{\beta\beta}$.
 
 We have also estimated the background from elastic neutrino-electron scatters from $^8$B solar neutrinos. As the endpoint of the electron recoil energy spectrum extends up to about 14\,MeV, such single-site scatters are a potential background source for double beta experiments. Using the $^8$B  neutrino flux of $\phi_{^8B}$=5.82$\times$10$^{6}$\,cm$^{-2}$s$^{-1}$ \cite{bahc05} and mean scattering cross sections of  $\sigma_{\nu_{e}}$=59.4$\times$10$^{-45}$\,cm$^2$ and $\sigma_{\nu_{\mu}}$=10.6$\times$10$^{-45}$\,cm$^2$ for electron- and muon-neutrinos respectively \cite{bahc89}, we obtain an event rate 0.036\,events/(t\,y) in the energy region of interest, see also Table~\ref{tab:bbrates}. While this is above the expected background from 2$\nu\beta\beta$-decays, and similar to the radon contribution, it is well below the one from detector materials. 
 
The expected total background in a $\pm$3$\sigma$ region around the $Q-$value is 4.6 events/(t$\cdot$y) in  6\,t of LXe fiducial mass. With an exposure of  30\,t$\cdot$y, a sensitivity to the neutrinoless double beta decay of T$_{1/2}^{0\nu}>$ 5.6$\times$10$^{26}$\,y (at 95\% C.L.), assuming a signal-to-background ratio of 1 and a 90\% detection efficiency, could be reached. This is more than a factor of 20 improvement with respect to the current lower limit on the half-life by EXO-200 \cite{Auger:2012ar} and KamLAND-Zen \cite{Gando:2012zm}. The sensitivity to the effective  Majorana neutrino mass would be in the range 20-60\,meV, depending on the value of the nuclear matrix element \cite{GomezCadenas:2011it}, allowing us to test part of the inverted-hierarchy region for neutrino masses.

In Figure \ref{fig:doublebeta}, right, we show the energy spectrum in the energy region of interest for the double beta decay. The background from detector components, from $^{222}$Rn and 2$\nu\beta\beta$-decays inside the liquid xenon, as well as from elastic $^{8}$B neutrino-electron scatters is displayed along with a hypothetical signal for an assumed half-life of 1.6$\times$10$^{25}$\,y.  Such a signal would provide, after 1 year of data taking,  106 events in a $\pm$3$\sigma$ region around the $Q-$value, with a background expectation of 28 events. The signal-to-background ratio can be enhanced by an improved energy resolution and by optimizing the energy window for the search, as the locations of the main background peak from $^{214}$Bi decays, at 2448\,keV, and the expected signal peak are shifted in energy with respect to one another. In Table \ref{tab:bbrates} we summarize the hypothetical signal and the expected backgrounds in the energy region of interest for the double beta decay.

\begin{table}[h!]
\centering
\begin{tabular}{|lc|}
\hline
                         &  Events/(6\,t$\cdot$y)    \\
Energy range & 2385--2533\,keV                \\
 \hline
Signal events & 106  \\
%\hline
Detector components & 27\\
$^{222}$Rn in LXe (0.1$\mu$Bq/kg) & 0.21\\
$^{8}$B ($\nu$-e scattering)& 0.22\\
$^{136}$Xe ($2\nu\beta\beta$)& 0.14\\
\hline
\end{tabular}
\caption{\small Expected signal for the neutrinoless double beta decay (for T$_{1/2}$ = 1.6$\times$10$^{25}$\,y) and background events from detector components, $^{222}$Rn in liquid xenon, $^{8}$B neutrino scatters and from $2\nu\beta\beta$ decays of $^{136}$Xe in the energy region 2385--2533\,keV. A fiducial mass of 6\,t of LXe and an energy resolution of $\sigma$/E = 1\% at the $Q-$value is assumed.}
\label{tab:bbrates}
\end{table}

An 'ultimate' sensitivity can be estimated assuming that the detector materials induced background can be reduced to negligible levels. In this case, the background from radon, neutrinos and the two neutrino double beta decay would yield an integrated rate of about 0.1\,events/(\,t$\cdot$y). Requiring once again a signal-to-background ratio of 1 would allow us to probe the half life of the neutrinoless mode up to T$_{1/2}$ = 8.5$\times$10$^{27}$\,y (at 95\% C.L.), after 10 years of data and using 14\,t of natural xenon.

\section{Summary and discussion}

We have studied the sensitivity of xenon-based, next-generation dark matter detectors to solar neutrinos, to coherent neutrino-nucleus scattering and to the neutrinoless double beta decay of $^{136}$Xe. As a concrete example, we considered the proposed DARWIN noble liquid TPC,  for which we made realistic assumptions used for a detailed background study. The xenon part of DARWIN, if realized with 21\,t (14\,t) of total (fiducial) liquid xenon mass, would observe, possibly for the first time, the real-time pp solar neutrino flux  via elastic neutrino-electron interactions. A rate of  about 1180\,events/y is to be expected for a realistic energy threshold of 2\,keV for electronic recoils and an upper energy boundary of 30\,keV. This measurement would improve the precision of the measured solar neutrino luminosity from 10\% to 1\%, after about  5 years of data, and reach the precision of theoretically predicted pp-neutrino rates when the solar photon luminosity is used as a constraint.  Electronic recoils from solar pp-neutrinos will limit the spin-independent dark matter sensitivity to a WIMP-nucleon cross section around 2$\times$10$^{-48}$cm$^2$, assuming that the remaining electronic recoil background can be reduced by a factor of 200, based on the observed  charge-to-light ratio  in the TPC. Nuclear recoils from coherent scattering of solar $^8$B neutrinos will limit the sensitivity to WIMP masses below $\sim$6\,GeV/c$^2$ to cross sections above $\sim$4$\times$10$^{-45}$cm$^2$, while higher energy, atmospheric neutrinos will challenge higher mass WIMP detection only below a cross section $\sim$10$^{-48}$cm$^2$. If an energy threshold based on the charge signal alone, as demonstrated by XENON10, is used in the data analysis, an overall rate of 532 coherent neutrino scattering events per year from solar neutrinos would be observed in 14\,t of LXe. 
A sensitivity to the neutrinoless double beta decay of $^{136}$Xe of T$_{1/2}^{0\nu}>$5.6$\times$10$^{26}$\,y (at 95\% C.L.) could be reached after 5 years of data taking, assuming a central mass of 6\,t of natural xenon, a 90\% signal detection efficiency, and a 99.8\% efficiency for tagging radon-induced backgrounds via the $^{214}$Bi-$^{214}$Po coincidence. An effective  Majorana neutrino mass in the range 20-60\,meV, depending on the value of the nuclear matrix element, could be probed, allowing us to test part of the inverted-hierarchy region for neutrino masses. This sensitivity could be drastically improved by using photosensors and a cryostat material with lower levels of $^{226}$Ra and $^{228}$Th than assumed in this study, for these components dominates the background around the double beta region of interest. 

In conclusion, {\sc DARWIN} would provide a unique opportunity not only to detect or exclude WIMPs with cross sections on nucleons as low as $\sim$10$^{-48}$cm$^2$, but also to precisely measure the low-energy solar neutrino spectrum and possibly to observe coherent neutrino-nucleus interactions from solar neutrinos, and the neutrinoless double beta decay of $^{136}$Xe.

\acknowledgments

This work is supported the Aspera first common call,  by the SNF grant 20AS21-129329 and by ITN Invisibles (Marie Curie Actions, PITN-GA-2011-289442). We would like to thank Carlos Pena-Garay for helpful discussions and for providing the numerical values for the curves in Figure 1, right.

\end{document}